\documentclass[aps,twocolumn,preprintnumbers,amsmath,amssymb,nofootinbib,superscriptaddress,notitlepage]{revtex4-1}
\usepackage{slashed}
\usepackage{soul}
\usepackage{mathrsfs,bm}
\usepackage{txfonts}
\usepackage{amssymb}
\usepackage{indentfirst}
\usepackage{graphicx,booktabs}
\usepackage{multirow}
\usepackage{overpic}
\usepackage{color}
\usepackage{amssymb}
\usepackage{epsfig}

\usepackage[utf8]{inputenc}

\begin{document}
\title{Rethinking the $P_c(4457)^+$ as the $P_{\psi}^{\Delta^+}(4457)$ isoquartet $\bar{D}^* \Sigma_c$ molecule}

\author{Fang-Zheng Peng}
\affiliation{School of Physics,  Beihang University, Beijing 100191, China}

\author{Mao-Jun Yan}
\affiliation{CAS Key Laboratory of Theoretical Physics, 
  Institute of Theoretical Physics,
  Chinese Academy of Sciences, Beijing 100190}

\author{Mario {S\'anchez S\'anchez}}
\affiliation{LP2IB (CNRS/IN2P3 – Universit\'e de Bordeaux), 33175 Gradignan cedex, France}
\affiliation{Departamento de Física, Universidad de Murcia, 30071 Murcia, Spain}

\author{Manuel {Pavon} Valderrama}\email{mpavon@buaa.edu.cn}
\affiliation{School of Physics,  Beihang University, Beijing 100191, China}

\date{\today}
\begin{abstract}
  The nature of the $P_{c}(4312)$, $P_c(4440)$ and $P_c(4457)$ pentaquarks
  is a fascinating theoretical question.
  Within the molecular picture their more usual interpretation is that of
  $I=\tfrac{1}{2}$ $\bar{D} \Sigma_c$ and $\bar{D}^* \Sigma_c$ bound
  states.
  Here we argue in favor of interpreting the $P_c(4457)$ pentaquark as a
  $I=\tfrac{3}{2}$ $\bar{D}^* \Sigma_c$ bound state
  (with spin $J=\tfrac{1}{2}$) instead.
  Owing to isospin symmetry breaking effects, with this identification
  the partial decay width of the $P_c(4457)^+$ into $J/\psi p$ will
  be of the same order of magnitude as the $P_{c}(4312)^+$ and
  $P_c(4440)^+$, in contrast with the considerably larger
  partial decay width in the $I=\tfrac{1}{2}$ scenario.
  In turn, this leads to a different hidden-charm molecular pentaquark spectrum,
  in which there are only four or five $P_{\psi}^N$ bound states instead of
  the usual seven, which might explain why the predicted $J=\tfrac{1}{2}$ and
  $\tfrac{3}{2}$ ($I=\tfrac{1}{2}$) $\bar{D}^* \Sigma_c^*$ molecular partners of
  the $P_c(4312)$ and $P_c(4440)$ have not been observed.
\end{abstract}

\maketitle

%\section{Introduction}

Four years ago the LHCb collaboration announced the discovery of three
hidden-charmed pentaquarks~\cite{Aaij:2019vzc} in the $J/\psi p$ invariant
mass distribution --- the $P_{c}(4312)$, $P_{c}(4440)$ and $P_{c}(4457)$ ---
with masses and widths (in units of ${\rm MeV}$)
\begin{eqnarray}
  M &=&4311.9\pm 0.7^{+6.8}_{-0.6} \, ,  \quad
  \Gamma =9.8\pm2.7^{+3.7}_{-4.5} \, , \label{eq:m1} \\
  M &=&4440.3\pm 1.3^{+4.1}_{-4.7} \, , \quad
  \Gamma =20.6\pm4.9^{+8.7}_{-10.1} \, , \label{eq:m2} \\
  M &=&4457.3\pm 0.6{}^{+4.1}_{-1.7} \, , \quad
  \Gamma =6.4\pm2.0^{+5.7}_{-1.9} \, , \label{eq:m3}
\end{eqnarray}
which we will refer to as $P_{c1}$, $P_{c2}$ and $P_{c3}$.
Their closeness to the $\bar{D} \Sigma_c$ and $\bar{D}^* \Sigma_c$ threshold,
together with the existence of previous predictions~\cite{Wu:2010jy,Wu:2010vk,Wu:2010rv,Xiao:2013yca}, has prompted their explanation as meson-baryon
bound states with $I=\tfrac{1}{2}$~\cite{Chen:2019bip,Chen:2019asm,Liu:2019tjn,Xiao:2019aya,Valderrama:2019chc,Liu:2019zvb,Guo:2019fdo},
though their nature is still far from determined
and there are alternative explanations too~\cite{Eides:2019tgv,Cheng:2019obk,Stancu:2020paw,Ferretti:2020ewe}.

Here we will consider the description of these three pentaquarks
in the molecular picture.
With the recent proposal of a new naming convention~\cite{Gershon:2022xnn},
the $P_{c1}$, $P_{c2}$ and $P_{c3}$ would be referred to as the $P_{\psi}^N(4312)$,
$P_{\psi}^N(4440)$ and $P_{\psi}^N(4457)$ within most of the molecular
interpretations available, where the superscript $N$ indicates
that in principle these pentaquarks are suspected to have
the same quantum numbers as a nucleon.
We will revisit this assumption for the case of the $P_c(4457)$, which
we argue is better explained if its quantum numbers are those of
the $\Delta$ isobar instead of the nucleon.
Thus it might be better referred to as the $P_{\psi}^{\Delta}(4457)$.

The usual molecular interpretation of the three LHCb pentaquarks
as $I=\tfrac{1}{2}$ (i.e. the octet representation
of SU(3)-flavor) $\bar{D} \Sigma_c$ and
$\bar{D} \Sigma_c^*$ states is not entirely free of problems.
As pointed out previously~\cite{Liu:2019tjn,Xiao:2019aya,Liu:2019zvb,Valderrama:2019chc},
this interpretation usually implies the existence of a heavy-quark spin
symmetry (HQSS) multiplet of seven molecular pentaquarks
where all possible octet $\bar{D}^{(*)} \Sigma_c^{(*)}$ configurations bind.
However, three of the pentaquarks in this multiplet will have markedly larger
partial decay widths into $J/\psi p$ than the others: the $J=\tfrac{1}{2}$
$\bar{D}^* \Sigma_c$ and the $J=\tfrac{1}{2}$ and $\tfrac{3}{2}$
$\bar{D}^* \Sigma_c^*$ states~\cite{Sakai:2019qph}.
Naively this implies that these last two $\bar{D}^* \Sigma_c^*$ states should
appear as prominent peaks in the $J/\psi p$ invariant mass distribution,
yet they don't, though this could be explained if their production
rates were to be smaller than those of the other pentaquarks.

It has also been noticed that within the molecular picture the interpretation
of the $P_c(4457)$ as an octet $\bar{D}^* \Sigma_c$ molecule is potentially
problematic.
In~\cite{Kuang:2020bnk} it is argued that the amplitude analysis of
the $\Lambda_b \to J/\psi p K^-$ decays suggests
the interpretation of $P_{c3}$ as a cusp rather
than as a bound state.
Ref.~\cite{Burns:2021jlu} considers the experimental constrains
from $\Lambda_b$ decays and photoproduction, which suggests that
while the $P_{c1}$ and $P_{c2}$ are easily explainable as $I=\tfrac{1}{2}$
$\bar{D} \Sigma_c$ and $J=\tfrac{3}{2}$ $\bar{D} \Sigma_c^*$
molecules, this is not the case for the $P_{c3}$.
More recently Ref.~\cite{Burns:2022uiv} argues from a fit to the $J/\psi p$
invariant spectrum for the interpretation of the $P_{c3}$ either as a
$\bar{D}^* \Sigma_c$ cusp, a $\bar{D} \Lambda_c(2595)$ triangular
singularity or a $\bar{D} \Lambda_c(2595)$ bound state.
This last interpretation has previously appeared in works about
the spectroscopy of molecular pentaquarks~\cite{Burns:2019iih,Peng:2020gwk}.

Along the present manuscript we will follow~\cite{Burns:2019iih,Peng:2020gwk}
and consider the $P_{c1}$ and $P_{c2}$ to be octet $J=\tfrac{1}{2}$
$\bar{D} \Sigma_c$ and $J=\tfrac{3}{2}$
$\bar{D} \Sigma_c^*$ bound states.
The opposite identification, namely $P_{c2}$ as a $J=\tfrac{1}{2}$
$\bar{D} \Sigma_c^*$ bound state, is more difficult to reconcile
with the known experimental information about this resonance
and it will thus not be considered in this work.
Our argument exploits isospin breaking effects to reduce the problematically
large $J/\psi p$ partial decay width of the $J=\tfrac{1}{2}$
$\bar{D} \Sigma_c^*$ configuration, where this mechanism only works
if we are dealing with a state that is close to threshold.
Previously isospin breaking effects have been discussed in the context of
the possible $P_c \to J/\psi \Delta$ decays of the $P_c(4457)$
as an octet pentaquark~\cite{Guo:2019fdo}.

Here we will explore a molecular explanation 
in which the $P_c(4457)$ or $P_{c3}$ is a decuplet ($I=\tfrac{3}{2}$)
$J=\tfrac{1}{2}$ $\bar{D} \Sigma_c^*$ molecule.
To illustrate the potential problems of the octet or $I=\tfrac{1}{2}$ molecular
description of the $P_c(4457)$ or $P_{c3}$ pentaquark,
we will begin by reviewing the decays of an $I=\tfrac{1}{2}$
$\bar{D}^{(*)} \Sigma_c^{(*)}$ meson-baryon pair into $J/\psi N$,
which can be derived from the light- and heavy-quark spin
decomposition of the meson-baryon pair~\cite{Sakai:2019qph},
yielding
\begin{eqnarray}
  \langle \bar{D} \Sigma_c(J=\tfrac{1}{2},I=\tfrac{1}{2}) | H | J/\psi N \rangle &=&
  \frac{1}{2\sqrt{3}}\,g  \, , \label{eq:amp-cc-Pc1} \\
  \langle \bar{D}^* \Sigma_c(J=\tfrac{1}{2},I=\tfrac{1}{2}) | H | J/\psi N \rangle &=&
  \frac{5}{6}\,g  \, , \label{eq:amp-cc-Pc2} \\
  \langle \bar{D}^* \Sigma_c(J=\tfrac{3}{2},I=\tfrac{1}{2}) | H | J/\psi N \rangle &=&
  -\frac{1}{3}\,g  \, , \label{eq:amp-cc-Pc3} 
\end{eqnarray}
with $g$ an unknown coupling constant. From the previous it is apparent that
the $J=\tfrac{1}{2}$ $\bar{D}^* \Sigma_c$ configuration has a particularly
large relative coupling with $J/\psi N$.
The partial decay width of a bound meson-baryon pair into $J/\psi N$ will
be given by
\begin{eqnarray}
  \Gamma(P_c \to J/\psi N) = \frac{p_{J/\psi N}}{\pi}\,\frac{\omega_{J/\psi} \omega_N}{m_{P_c}}\,g_{P_c}^2\,
  {\left| \Psi_{P_c}(0) \right|}^2 \, ,
\end{eqnarray}
with $p_{J/\psi N}$ the center-of-mass momentum of
the final $J/\psi N$ state, $\omega_{J/\psi}$ and $\omega_N$
the energies of the final $J/\psi$ and $N$, $m_{P_c}$ the pentaquark mass,
$g_{P_c}$ the coupling $g$ times the numerical
factor from the light- and heavy-quark spin decomposition
in Eqs.(\ref{eq:amp-cc-Pc1}-\ref{eq:amp-cc-Pc3}),
and $\Psi_{P_c}(0)$ the r-space wave function of
the pentaquark at the origin ($\vec{r} = 0$).
If we assume that the pentaquarks can be described in a contact-range theory,
the r-space wave function at the origin takes the form
\begin{eqnarray}
  \Psi_{P_c}(0) = \mathcal{N}_{P_c}\, \int \frac{d^3 \vec{q}}{(2 \pi)^2}\,
      \frac{f(q/\Lambda)}{{2 \mu_{P_c} B_{P_c}} + {\vec{q}\,}^2} \, ,
\end{eqnarray}
where $\mathcal{N}_{P_c}$ is the normalization of the wave function,
$f(x)$ a regulator function, $\mu_{P_c}$ the reduced mass of
the meson-baryon system and $B_{P_c}$ the binding energy.
If we use a Gaussian regulator $f(x) = e^{-x^2}$ and a cutoff
$\Lambda = 0.75\,{\rm GeV}$ (of the order of the $\rho$ meson mass~\footnote{This
  value of the cutoff maximizes the momenta for which the contact-range
  description is valid ($k < \Lambda$, with $k$ the center-of-mass momentum
  of the meson-baryon system), while not being as hard as to
  resolve the short-range details of the meson-baryon potential
  ($\Lambda > m_{\rho}$ with $m_{\rho}$ the $\rho$ mass, if we assume
  that their short-range potential is described
  by vector meson exchange).}),
the ratio of the partial decay widths for
$P_{c1}$, $P_{c2}$ and $P_{c3}$ will be
\begin{eqnarray}
 % \frac{{\tilde{\Gamma}_i}}{{\tilde{\Gamma}_1}} =
  1 : 1.8 : 11.5 \, , \label{eq:JPsi-octet}
\end{eqnarray}
respectively, where it can be appreciated that the $J=\tfrac{1}{2}$ $P_{c3}$
pentaquark has a $J/\psi N$ partial decay width one order of
magnitude larger than the other two pentaquarks,
as previously discussed in~\cite{Sakai:2019qph,Burns:2021jlu}.

Experimentally what we know are the {\it production fractions}~\footnote{The
  quantity that we name here as production fraction is actually closely
  related to the fit fraction --- loosely speaking, the fraction of
  the $X \to ABC$ decay that has a resonance $R$ as an intermediate state,
  i.e. $X \to A(R) \to ABC$ --- though they are not equivalent.
  For a detailed discussion on their relation,
  we recommend Ref.~\cite{Burns:2021jlu}.
}
of each of the pentaquarks, defined as
\begin{eqnarray}
  {\mathcal F}(P_c) = \frac{\mathcal{B}(\Lambda_b^0 \to K^- P_c^+)\,\mathcal{B}(P_c^+ \to J/\psi p)}{\mathcal{B}(\Lambda_b^0 \to K^- J/\psi p)} \, ,
  \label{eq:branching-fraction}
\end{eqnarray}
where $\mathcal{B}$ denotes the branching ratio of a particular decay,
with ${\mathcal F}_{i} = 0.30^{+0.35}_{-0.11}$, $1.11^{+0.40}_{-0.34}$,
$0.53^{+0.22}_{-0.21}$ for $i = 1,2,3$~\cite{Aaij:2019vzc}.
The ratios of the production fractions are then % $1:3.7:1.8$,
\begin{eqnarray}
  \frac{{\mathcal F}_{i}}{{\mathcal F}_{1}}\Big|_{\rm exp} =
  1 : 3.7^{+2.5}_{-2.3} : (1.8 \pm 1.2) \, , \label{eq:fit-ratios-exp}
\end{eqnarray}
which though not directly
comparable with the $J/\psi p$ partial decay width ratios of
Eq.~(\ref{eq:JPsi-octet})  should still be of
the same order of magnitude.
Actually, only the ratios of $\mathcal{B}(\Lambda_b^0 \to K^- P_{ci}^+)$
are missing for a full comparison (the ratios of the
$\mathcal{B}(P_{ci}^+ \to J/\psi p)$ can be obtained from the experimental
decay widths and the relative partial decay widths
in Eq.~(\ref{eq:JPsi-octet})).
If we define $\mathcal{R}_i = \mathcal{B}(\Lambda_b^0 \to K^- P_{ci}^+) / \mathcal{B}(\Lambda_b^0 \to K^- P_c^+(4312))$ with $i= 2,3$, we can express the
theoretical prediction for the production fractions as
\begin{eqnarray}
  \frac{{\mathcal F}_{i}}{{\mathcal F}_{1}}\Big|_{\rm th} =
  1 : (0.86^{+1.10}_{-0.53})\,\mathcal{R}_2: (18^{+16}_{-13})\,{\mathcal R}_3 \, ,
\end{eqnarray}
which also includes the uncertainties coming from the experimental
decay widths (summed in quadrature).
These ratios indicate that unless the relative production rate ${\mathcal R}_3$
for the $P_c(4457)$ pentaquark is considerably smaller
than for the other two, we will have an inconsistency
with the experimental data.

\begin{figure}
  \begin{center}
    \epsfig{figure=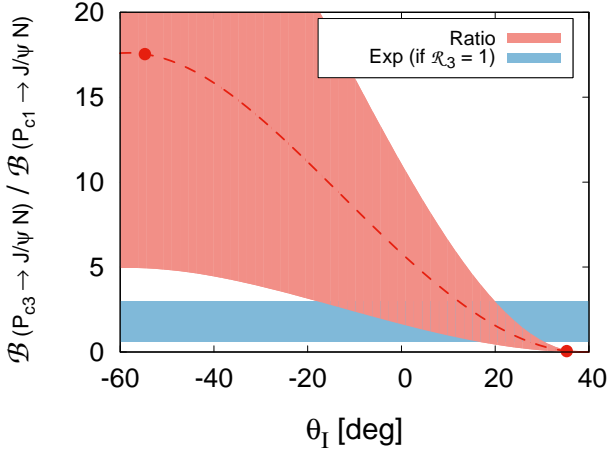,
      width=8.5cm}
    \end{center}
  \caption{
    Dependence on the isospin angle of the ratio of the branching ratios of
    the $P_c(4457)$ ($P_{c3}$) and $P_c(4312)$ ($P_{c1}$) decays into $J/\psi N$.
    That is, we plot $\mathcal{B}_3 / \mathcal{B}_1$ as a function of
    $\theta_I$ (as defined in Eq.~(\ref{eq:isospin})), where
    $\mathcal{B}_i = \mathcal{B}(P_{ci} \to J/\psi N) =
    {\Gamma}(P_{ci} \to J/\psi N) / \Gamma(P_{ci})$.
    This is related to the ratio of production fractions of
    Eq.~(\ref{eq:fit-ratios-exp}) by $\mathcal{B}_3 / \mathcal{B}_1 = \mathcal{F}_3 / (\mathcal{F}_1 \mathcal{R}_3$), with $\mathcal{F}_i$ the production
    fraction of pentaquark $i$ (check Eq.~(\ref{eq:branching-fraction})) and
    $\mathcal{R}_3 = \mathcal{B}(\Lambda_b^0 \to K^- P_{c}^+(4457)) / \mathcal{B}(\Lambda_b^0 \to K^- P_c^+(4312))$
    the relative production rate of the $P_c(4457)$
    with respect to the $P_c(4312)$.
    The dashed line represents the central value, which is almost independent of
    the choice of cutoff or the uncertainties in the binding energies of
    the pentaquarks; the two small, solid circles represent
    the values for $I=1/2$ ($\theta_I = -54.7^{\circ}$) and
    $I=3/2$ ($\theta_I = +35.3^{\circ}$), respectively;
    the band around the dashed line
    (labeled ``Ratio'') represents the uncertainty coming
    from the ratio of the experimental widths of
    the pentaquarks ($\Gamma_1 / \Gamma_3 = 1.53^{+1.39}_{-1.10}$).
    The horizontal band (labeled ``Exp'') represents the experimental
    $\mathcal{B}_3 / \mathcal{B}_1$ ratio under the assumption that
    the production rates are identical, i.e. $\mathcal{R}_3 = 1$.
    $\mathcal{R}_3 > 1$ ($\mathcal{R}_3 < 1$) will favor the interpretation of
    the $P_c(4457)$ as an $I=3/2$ isoquartet ($I=1/2$ isodoublet)
    $\bar{D}^* \Sigma_c$ bound state.
  }
\label{fig:branching}
\end{figure}

This changes if we consider the $P_{c3}$ isospin wave function
\begin{eqnarray}
  | P_c(4457)^+ \rangle = \cos{\theta_I} \, | \bar{D}^{*0} \Sigma_c^+ \rangle
  + \sin{\theta_I}\,|D^{*-} \Sigma_c^{++} \rangle \, , \label{eq:isospin}
\end{eqnarray}
with the isospin angle $\theta_I = -54.7^{\circ}$ and $35.3^{\circ}$ for a pure
$I=\tfrac{1}{2}$ and $\tfrac{3}{2}$ state, respectively.
If we additionally assume a $J=\tfrac{1}{2}$ $P_{c3}$, the decay amplitude reads
\begin{eqnarray}
  && \langle \bar{D}^* \Sigma_c(J=\tfrac{1}{2}) | H | J/\psi p \rangle = \nonumber \\ && \qquad \qquad
  \frac{5}{6}\,g\,(\frac{1}{\sqrt{3}}\cos{\theta_I} - \sqrt{\frac{2}{3}}\sin{\theta_I})  \, . \label{eq:amp-cc-Pc12-isospin} 
\end{eqnarray}
When including the effects coming from the different masses of the
$\bar{D}^{*0}\Sigma_c^+$ and ${D}^{*-}\Sigma_c^{++}$ thresholds,
we find the partial decay width ratios
\begin{eqnarray}
  1 : 1.8 : 1.0 \quad \mbox{or} \quad 1:1.8:0.035 \, ,
\end{eqnarray}
for $\theta_I = 20.1^{\circ}$ and $35.3^{\circ}$, respectively.
Alternatively, if we consider instead the ratios of the production
fractions we find
\begin{eqnarray}
  \frac{{\mathcal F}_{i}}{{\mathcal F}_{1}}\Big|_{\rm th} = \,\,
  && 1 : (0.86^{+1.10}_{-0.53})\,\mathcal{R}_2 : (1.5^{+1.4}_{-1.1})\,\mathcal{R}_3 \quad \mbox{or} \quad \nonumber \\
  && 1 : (0.86^{+1.10}_{-0.53})\,\mathcal{R}_2 : (0.054^{+0.049}_{-0.042})\,\mathcal{R}_3 \, ,
\end{eqnarray}
for $\theta_I = 20.1^{\circ}$ and $35.3^{\circ}$, with $\mathcal{R}_i$
the ratios of $\mathcal{B}(\Lambda_b^0 \to K^- P_{ci}^+)$.
In Fig.~\ref{fig:branching} we illustrate the dependence of
${\mathcal F}_{3}/ ({\mathcal F}_{1} \mathcal{R}_3)$
on the isospin angle $\theta_I$ and compare it with the experimental  
${\mathcal F}_{3}/ {\mathcal F}_{1} |_{\rm exp} = 1.8 \pm 1.2$
under the assumption that the production rates are identical
($\mathcal{R}_3 = 1$).
From Fig.~\ref{fig:branching} it can be appreciated that even for identical
production rates there is a wide band of values of $\theta_I$ that
are potentially compatible with the experimental
production fractions, though with a preference
for values of $\theta_I$ closer to $I=3/2$ than to $I=1/2$
(unless $\mathcal{R}_3 \ll 1$).

A second argument in favor of the $P_{\psi}^{\Delta}(4457)$ assignment comes
from the full decay widths of the pentaquarks, which are thought to be
dominated by the meson-baryon decays, i.e. the $\bar{D} \Lambda_c$ and
$\bar{D}^{*}\Lambda_c$ channels.
The reason is (a) the very small decay branching ratios of the pentaquarks
into $J/\psi p$, which in the GlueX experiment~\cite{Ali:2019lzf}
have been found to be $\mathcal{B}(P_{ci} \to J/\psi p) < 4.6, 2.3, 3.8\,\%$
for $i=1,2,3$, respectively, (b) the relatively small partial decay
width into $\bar{D}^{(*)} \Lambda_c \pi$ of the order
of $2\,{\rm MeV}$~\cite{Burns:2021jlu} (i.e. about the width of
a free $\Sigma_c$~\cite{Voloshin:2019aut}).
The decay amplitudes into $\bar{D}^{(*)} \Lambda_c$
are given by~\cite{Du:2021fmf,Yan:2021nio,Burns:2022uiv}
\begin{eqnarray}
  \langle \bar{D} \Sigma_c(J=\tfrac{1}{2},I=\tfrac{1}{2}) | H | \bar{D}^* \Lambda_c \rangle &=&
  \sqrt{3}\,E_b  \, , \label{eq:amp-mol-Pc1} \\
  \langle \bar{D}^* \Sigma_c(J=\tfrac{1}{2},I=\tfrac{1}{2}) | H | \bar{D} \Lambda_c \rangle &=&
  \sqrt{3} E_b \, , \label{eq:amp-mol-Pc2} \\
  \langle \bar{D}^* \Sigma_c(J=\tfrac{1}{2},I=\tfrac{1}{2}) | H | \bar{D}^* \Lambda_c \rangle &=&
  -2 E_b \, , \label{eq:amp-mol-Pc3} \\
  \langle \bar{D}^* \Sigma_c(J=\tfrac{3}{2},I=\tfrac{1}{2}) | H | \bar{D}^* \Lambda_c \rangle &=&
  E_b  \, , \label{eq:amp-mol-Pc4} 
\end{eqnarray}
which, if we determine the coupling $E_b$ from the $P_c(4312)$ decay width,
will lead to 
\begin{eqnarray}
  \Gamma_{P_{c2}} \simeq 9.8^{+4.6}_{-5.2} \,{\rm MeV} \quad \mbox{and} \quad \Gamma_{P_{c3}} \simeq 81^{+38}_{-43} \, {\rm MeV} \, .
\end{eqnarray}
In the case of the $P_{c3}$ the calculated width is clearly inconsistent
with the experimental one.
In this case, the assumption of a predominantly $I=\tfrac{3}{2}$ $P_{c3}$ plus
isospin breaking effects will lead instead to
\begin{eqnarray}
  \Gamma_{P_{c3}} \simeq  7.0^{+3.3}_{-3.7}\,{\rm MeV} \quad \mbox{or} \quad 0.2 \pm 0.1\,{\rm MeV} \, ,
\end{eqnarray}
for $\theta_I = 20.1^{\circ}$ and $35.3^{\circ}$, respectively.
This comparison is however considerably less reliable than the $J/\psi N$ one
for the following reasons:
\begin{itemize}
\item[(i)]
First, if the pentaquarks are molecular, being as close as they are to
threshold, the Breit-Wigner resonance profile might not be ideal,
meaning that their actual widths might be different
from the experimental ones~\cite{Fernandez-Ramirez:2019koa}.
This is in contrast with the $J/\psi N$ partial decay widths, which are
an important factor in how visible the pentaquarks are
in the $J/\psi N$ invariant mass distribution.
\item[(ii)]
Second, the extraction of the branching ratios for the $J / \psi p$
decays from GlueX~\cite{Ali:2019lzf} depends on the JPAC model
for $J / \psi$ photoproduction~\cite{HillerBlin:2016odx}.
  In view of the recent GlueX photoproduction data~\cite{GlueX:2023pev},
  the JPAC collaboration itself has updated its priors regarding one of
  the assumptions within their model --- vector meson dominance
  in the heavy sector --- which is no longer considered
  to be reliable~\cite{JointPhysicsAnalysisCenter:2023qgg}.
  As a consequence the actual $J / \psi p$ branching ratios 
  might be very different from current estimations.
  In particular, if the $J / \psi p$ branching ratios happen to be much above
  the single digit percentage level estimated by GlueX~\cite{Ali:2019lzf},
  then the possible inconsistencies in the $\bar{D} \Lambda_c$ and
  $\bar{D}^* \Lambda_c$ decays would be a secondary concern (though
  in this case the $J / \psi p$ partial decay widths will represent
  a larger contribution of the pentaquark widths, and the problems
  related with them will be less dependent on the unknown
  production rates and thus more pressing).
  Yet, the fit of Ref.~\cite{Burns:2022uiv} to the $J/\psi p$
  invariant spectrum, which does not rely on vector meson dominance,
  suggests even smaller $J / \psi p$ branching ratios (of the order
  of $10^{-3}$) than those of GlueX.
  Be it as it may, future experimental and theoretical results
  will be necessary to better determine these ratios.
\item[(iii)] Third, 
very probably the final meson-baryon states are strongly interacting,
which might in turn change considerably the previous predictions
for the decay widths.
This is particularly true for the $P_c(4312) \to \bar{D}^* \Lambda_c$ decay,
in which the final center-of-mass momentum of the meson-baryon system
is merely $190\,{\rm MeV}$, i.e. not that far from threshold,
suggesting the possibility of a considerably larger partial decay width
it the $\bar{D}^* \Lambda_c$ interaction is attractive.
\item[(iv)] Fourth,
the center-of-mass momentum is very different for the three pentaquarks,
from which we do not only expect a different role of the final state interaction
but also of the relative importance of finite hadron size effects
(e.g. form factors and regulators).
\item[(v)] Fifth, we have assumed that these decays are S-wave. Yet, it could happen
that D-wave decays are as important or more than the S-wave ones.
\end{itemize}
For these reasons, even though the comparison of the $\bar{D}^{(*)} \Lambda_c$
decays is interesting, they should be considered as 
less compelling than the $J/\psi N$ ones.

Next we will consider the $P_c(4457)$ spectroscopy.
We will begin with the description of the molecular pentaquarks
within the lowest order (${\rm LO}$) of a contact-range EFT,
as has been discussed in the literature~\cite{Liu:2018zzu,Liu:2019tjn}.
Within this type of EFT the momentum space potentials
for the $I=\tfrac{1}{2}$ or octet configurations are
\begin{eqnarray}
  V(\bar{D} \Sigma_c, I=\tfrac{1}{2}) &=& C^O_a \, , \\
  V(\bar{D}^* \Sigma_c, J=\tfrac{1}{2}, I=\tfrac{1}{2}) &=& C^O_a - \frac{4}{3} C_b^O\, , \\
  V(\bar{D}^* \Sigma_c, J=\tfrac{1}{2}, I=\tfrac{1}{2}) &=& C^O_a + \frac{2}{3} C_b^O\, ,
\end{eqnarray}
with $C^O_a$ and $C^O_b$ the two octet couplings.
These potentials are singular (they correspond to a Dirac-delta in coordinate
space) and require regularization, for which we choose here a separable
regulator of the type
\begin{eqnarray}
  \langle p' | V_C | p \rangle = C(\Lambda)\,
  f(\frac{p}{\Lambda})\,f(\frac{p'}{\Lambda}) \, ,
\end{eqnarray}
where $C(\Lambda)$ is the coupling (which runs with the cutoff) and $f(x)$
a regulator function, for which we use a Gaussian: $f(x) = e^{-x^2}$.
Then, to find the poles, we plug this potential
into the Lippmann-Schwinger equation 
\begin{eqnarray}
  1 + 2\mu\,C(\Lambda)\,\int \frac{d^3 \vec{q}}{(2 \pi)^3}\,
  \frac{f^2(\frac{q}{\Lambda})}{M_{\rm th} + \frac{q^2}{2 \mu} - M_{\rm P}}
   = 0\, , \label{eq:bound}
\end{eqnarray}
with $\mu$ the reduced mass, $M_{\rm th}$ the threshold mass and $M_{\rm P}$
the mass of the molecular pentaquark.
The coupling $C(\Lambda)$ is renormalized from the condition of
reproducing $M_{\rm P}$ for a given molecular pentaquark
candidate.

The octet molecular pentaquark potential contains only two parameters but
there are three pentaquarks, which means that we can check
the consistency of the molecular hypothesis by using two
pentaquarks as input and predicting the third one.
If we assume that the $P_{c2}$ and $P_{c3}$ are spin $J=\tfrac{3}{2}$ and
$\tfrac{1}{2}$, respectively (in agreement
with the arguments of~\cite{Peng:2020xrf}
and previous explorations in the molecular model~\cite{Yamaguchi:2019seo,Valderrama:2019chc,Liu:2019zvb,Du:2019pij,Peng:2021hkr,Yamaguchi:2019seo}),
there are three possible ways in which to calibrate
the $C_a^O$ and $C_b^O$ couplings: (i) from the $P_{c1}$ and $P_{c2}$ pentaquarks,
(ii) from the $P_{c2}$ and $P_{c3}$ and (iii) from the $P_{c1}$ and $P_{c3}$.
We will refer to them as set (i), (ii) and (iii).

For the uncertainties, we will consider the following two error sources and
sum them in quadrature.
The first is varying the cutoff around the $\Lambda = 0.75\,{\rm GeV}$ central
value, for which we choose the $(0.5-1.0)\,{\rm GeV}$ window~\footnote{That is,
  we vary the cutoff around the momentum scale at which we expect the EFT
  description to break down (e.g. the $\rho$ meson mass).}.
The second is the truncation error of the couplings: $C_a^O$ and $C_b^O$ are
leading order couplings, yet the couplings that we know ($C(P_{ci})$ with
$i=1,2,3$) contain all subleading order corrections,
which are undetermined by the available
experimental data.
Thus the ${\rm LO}$ couplings might differ from the full ones by a relative
error of $\gamma / m_{\rho}$ (the EFT truncation error), with $\gamma$
the wave number of the pentaquark from which the coupling is obtained
and $m_{\rho}$ the rho meson mass (we remind that
$\gamma = \sqrt{2 \mu (M_{\rm th} - M_{\rm P})}$, with $\mu$, $M_{\rm th}$ and
$M_{\rm P}$ defined below Eq.~(\ref{eq:bound})).
For instance, if $C_a^O$ is the ${\rm LO}$ coupling and $C(P_{c1})$ the coupling
reproducing the $P_{c1}$ pentaquark, the uncertainty will be determined
from the condition $C_a^O\,(1 + \mathcal{O}(\gamma/m_{\rho}) ) = C(P_{c1})$.
With this for each set we get the couplings
(i) $C_a^O = -1.19^{+0.17}_{-0.26}$ ($-(2.16-0.80)$) ${\rm fm}^2$ and
$C_b^O = -0.38^{+0.08}_{-0.15}$ ($-(1.07-0.18)$) ${\rm fm}^2$,
(ii)  $C_a^O = -1.30^{+0.13}_{-0.19}$ ($-(2.52-0.85)$) ${\rm fm}^2$ and
$C_b^O = -0.207^{+0.046}_{-0.082}$ ($-(0.543-0.107)$) ${\rm fm}^2$,
(iii) $C_a^O = -1.19^{+0.17}_{-0.26}$ ($-(2.16-0.80)$) ${\rm fm}^2$ and
$C_b^O = -0.120^{+0.014}_{-0.018}$ ($-(0.279-0.068)$) ${\rm fm}^2$.
The first value is the $\Lambda = 0.75\,{\rm GeV}$ coupling and
its expected EFT truncation error, while the values in parentheses
represent their determination for $\Lambda = (0.5-1.0)\,{\rm GeV}$.
These uncertainties can then be propagated into
the predicted mass of the third pentaquark.

We show the results in Table \ref{tab:consistency}, where it can be appreciated
that in general this works well: the hypothesis that the three pentaquarks
are molecular is self-consistent.
However, if the input states are the $P_{c1}$ and $P_{c2}$,
the $P_c(4457)$ is predicted as a near threshold
virtual (instead of bound) state.
This by itself is not a serious issue, as a virtual state close to threshold
could still be detected in experiments.
But it will be interesting to explore the possibility that the $P_c(4457)$
is an $I=\tfrac{3}{2}$ state.

\begin{table}[t]
  \begin{center}
    \begin{tabular}{|cccccc|}
      \hline \hline
      Molecule & $I(J^P)$ & M(set i) & M(set ii) & M(set iii) & Experiment \\
      \hline \hline
      $\bar{D} \Sigma_c$ & $\frac{1}{2}(\frac{1}{2}^-)$ & Input & $4307.3^{+5.2}_{-8.8}$ & Input &
      $4311.9^{+6.8}_{-0.9}$ \\
      $\bar{D}^* \Sigma_c$ & $\frac{1}{2}(\frac{3}{2}^-)$ & Input & Input & $4448.5^{+7.0}_{-12.5}$ &
      $4440.3^{+4.3}_{-4.9}$ \\
      $\bar{D}^* \Sigma_c$ & $\frac{1}{2}(\frac{1}{2}^-)$ & ${(4461.7^{+0.3(B)}_{-8.6})}^V$ & Input & Input &
      $4457.3^{+4.1}_{-1.8}$ \\
      \hline \hline
    \end{tabular}
    \caption{Consistency of the $I=\tfrac{1}{2}$ hypothesis
      for the three pentaquarks observed in~\cite{Aaij:2019vzc}.
      ``Molecule'' indicates the meson-baryon system, $I(J^P)$ the isospin,
      spin and parity and $M$ the masses (in ${\rm MeV}$)
      in set i, ii and iii as defined
      in the main text, where a $V$ superscript indicates a virtual state
      and no superscript a bound state.
      We find that the three sets are consistent
      with the experimental masses of the pentaquarks.
    }
    \label{tab:consistency}
  \end{center}
\end{table}

\begin{table*}[htt]
  \begin{center}
    \begin{tabular}{|c|c|cc|cc|}
      \hline \hline
      Molecule & $M_{\rm th}$ & $I(J^P)$ & M & $I(J^P)$ & M  \\
      \hline \hline
      $\bar{D} \Sigma_c$ & $4320.7$ & $\frac{1}{2}(\frac{1}{2}^-)$ & Input &
      $\frac{3}{2}(\frac{1}{2}^-)$ & - \\
      $\bar{D} \Sigma_c^*$ & $4385.4$ & $\frac{1}{2}(\frac{3}{2}^-)$ & $4376.2^{+5.7}_{-11.2}$ &
      $\frac{3}{2}(\frac{3}{2}^-)$ & - \\
      $\bar{D}^{*0} \Sigma_c^{+}$-${D}^{*-} \Sigma_c^{++}$ &  $4459.5$, $4464.2$ &
      $\frac{1}{2}(\frac{1}{2}^-)$ & $4464.4^{+2.1}_{-6.7} - (0.0^{+4.2}_{-0.0}) \,i$ &
      $\frac{3}{2}(\frac{1}{2}^-)$ & Input \\
      $\bar{D}^* \Sigma_c$ & $4462.0$ & $\frac{1}{2}(\frac{3}{2}^-)$ & Input &
      $\frac{3}{2}(\frac{3}{2}^-)$ & - \\
      $\bar{D}^{*0} \Sigma_c^{*+}$-${D}^{*-} \Sigma_c^{*++}$ & $4524.3$, $4528.7$ &
      $\frac{1}{2}(\frac{1}{2}^-)$ &
      $4531.3^{+9.7}_{-2.6} - (2.7^{+10.2}_{-2.7})\, i$ & 
      $\frac{3}{2}(\frac{1}{2}^-)$ &  $4514^{+10}_{-39}$ \\
      $\bar{D}^{*0} \Sigma_c^{*+}$-${D}^{*-} \Sigma_c^{*++}$ & $4524.3$, $4528.7$ &
      $\frac{1}{2}(\frac{3}{2}^-)$ & $4523.8^{+0.5}_{-7.7}$ &
      $\frac{3}{2}(\frac{3}{2}^-)$ & $4530.3^{+7.0}_{-4.6} - (1.9_{-1.9}^{+10.2}) \,i$ \\
      $\bar{D}^* \Sigma_c^*$ & $4526.7$ & $\frac{1}{2}(\frac{5}{2}^-)$ & $4498^{+10}_{-17}$  &
      $\frac{3}{2}(\frac{5}{2}^-)$ &  - \\
      \hline \hline
    \end{tabular}
    \caption{Predictions for the octet ($I=\tfrac{1}{2}$) and decuplet
      ($I=\tfrac{3}{2}$) molecular pentaquark spectrum
      from the conditions of (a) reproducing the $P_c(4312)$ and $P_c(4440)$
      as $I=\tfrac{1}{2}$ $\bar{D} \Sigma_c$ and $J=\tfrac{3}{2}$
      $\bar{D}^* \Sigma_c$ bound states, (b) the $P_c(4457)$ as a
      $I=\tfrac{3}{2}$, $J=\tfrac{3}{2}$ $\bar{D}^* \Sigma_c$ molecule and
      (c) the phenomenologically inspired relation
      $C_b^D = - 2 C_b^O$.
      ``Molecule'' shows the meson-baryon configuration (when
      two configurations are shown it indicates that we include
      explicit isospin breaking effects, which we only do for pentaquarks
      predicted really close to threshold), $I(J^P)$ are the isospin, spin
      and parity of the state and $M$ is the mass (in MeV).
      For molecules in which we include isospin breaking the criterion
      for considering it as $I=\tfrac{1}{2}$ and $I=\tfrac{3}{2}$ is
      that the central value of the isospin angle is $\theta_I < 0$
      and $\theta_I > 0$, respectively (see Eq.~(\ref{eq:isospin})).
      However, owing to the large uncertainties in $\theta_I$, the opposite
      isospin identification cannot be completely discarded for
      the $J=\tfrac{1}{2}$ $\bar{D}^* \Sigma_c$ and
      $J=\tfrac{3}{2}$ $\bar{D}^* \Sigma_c^*$ molecules.
     }
    \label{tab:predictions}
  \end{center}
\end{table*}

For this last scenario --- $P_c(4312)$ and $P_c(4440)$ as octets and
$P_{c}(4457)$ as a decuplet --- we will explicitly include isospin
breaking effects for the $P_c(4457)$, in which case
the potential (in the $\bar{D}^{*0} \Sigma_c^{+}$-${D}^{*-} \Sigma_c^{++}$ basis)
will read 
\begin{eqnarray}
  V_C(\bar{D}^* \Sigma_c, J=\tfrac{1}{2}) =
  \begin{pmatrix}
    \frac{1}{3}\,C^O + \frac{2}{3}\,C^D & -\frac{\sqrt{2}}{3}
    \left( C^O - C^D \right) \\
    -\frac{\sqrt{2}}{3} \left( C^O - C^D \right) &
    \frac{2}{3}\,C^O + \frac{1}{3}\,C^D 
  \end{pmatrix} \, , \nonumber \\
\end{eqnarray}
where $C^O$ and $C^D$ are the octet and decuplet potential
for this configuration, i.e. $C^O = C^O_a - \frac{4}{3} C^O_b$
and $C^D = C^D_a - \frac{4}{3} C^D_b$.
If we calibrate $C^D$ as to reproduce the location of the $P_c(4457)$,
we obtain $C^D = -0.97^{+0.11}_{-0.14}\,$ ($-(1.65-0.70)$) ${\rm fm}^2$
and $\theta_I = {(26.6^{+7.6}_{-28.4})}^{\circ}\,({(30.7-23.3)}^{\circ})$, which is
relatively close to a pure $I=\tfrac{3}{2}$ state and where the errors
come from propagating the uncertainties of $C_a^O$ and $C_b^O$ into
$C^D$ and $\theta_I$.
From this angle, the relative partial decay widths into $J/\psi p$ of the
three pentaquarks will be
\begin{eqnarray}
  1 : 1.83\,(1.84-1.82) : 0.44^{+3.64}_{-0.41}\,(0.26-0.66) \, ,
\end{eqnarray}
to be compared with Eq.~(\ref{eq:JPsi-octet}) for the octet hypothesis.
The previous will translate into a production fraction ratio of
\begin{eqnarray}
  \frac{{\mathcal F}_{i}}{{\mathcal F}_{1}}\Big|_{\rm th} =
  1 &:& 0.87^{+1.10}_{-0.53}\,(0.88-0.87)\,\mathcal{R}_2 \nonumber \\
  &:& 0.67^{+5.69}_{-0.67}\,(0.40-1.01)\,{\mathcal R}_3 \, ,
\end{eqnarray}
which also include the uncertainties in the experimental decay widths.
These values are potentially more compatible with the experimental
production fractions of Eq.~(\ref{eq:fit-ratios-exp}) without
requiring unnaturally large or small ${\mathcal R}_{2}$ and ${\mathcal R}_{3}$.
Meanwhile the decay width into $\bar{D}^{(*)} \Lambda_c$ is estimated to be
$3.1^{+28.1}_{-3.1}\,(1.2-6.1)\,{\rm MeV}$,
which is of the correct order of magnitude.

From the previous it is in principle not possible to derive
the spectrum of the decuplet pentaquarks, as we do not know
the specific values of $C^D_a$ and $C^D_b$.
However, it is possible to make an educated guess on the basis that
the spin-spin contact-range coupling $C_b$ represents the spin-spin
dependence coming from the short range vector meson exchange
potential, which is in turn given by~\cite{Liu:2019zvb}
\begin{eqnarray}
  V_{Vb}(r) = \left(1 + \vec{\tau}_1 \cdot \vec{T}_2 \right)\,
  f_{V1} f_{V2}\,\frac{m_V^2}{6 M^2}\,\frac{e^{-m_V r}}{4 \pi r} \, ,
\end{eqnarray}
with $r$ the distance, $f_{V1}$ and $f_{V2}$ coupling constants, $m_V$ the
vector meson mass and $M$ a scaling mass (e.g. the nucleon mass).
The isospin factor in front of the potential takes the values
$(1 + \vec{\tau}_1 \cdot \vec{T}_2) = -1$, $2$
for $I=\tfrac{1}{2}$ and $\tfrac{3}{2}$, respectively.
That is, for the short-range vector exchange potential
from which $C_b$ is derived $V_{Vb}^D = - 2 V_{Vb}^O$.
We might simply assume this to be true for the contact coupling $C_b$ as well,
in which case $C_b^D = - 2 C_b^O \,(1 \pm \delta_{\rm V})$, with $\delta_{\rm V}$
the expected relative error for this relation,
which we will set to be $30\%$.
With this assumption we end up with
$C_a^D = 0.05^{+0.52}_{-0.40}$ ($(1.20-(-0.19))$) ${\rm fm}^2$ and
$C_b^D = 0.76^{+0.38}_{-0.28}$ ($(2.14-0.37)$) ${\rm fm}^2$.

From this we are able to derive the octet and decuplet pentaquark spectra,
which we show in Table \ref{tab:predictions}.
It is interesting to notice that we predict seven pentaquarks bound below
their respective thresholds, in agreement with most works
implementing HQSS constraints~\cite{Xiao:2013yca,Liu:2019tjn,Xiao:2019aya,Sakai:2019qph,Du:2019pij,Dong:2021juy,Du:2021fmf}.
But the quantum numbers are not the same, as the previous works usually
predict seven octet pentaquarks (while here there are five octet and
two decuplet bound pentaquarks, plus a few virtual states and
resonances).
In this regard our octet spectrum is more similar to the one we originally
proposed in~\cite{Liu:2018zzu} than to our later predictions~\cite{Liu:2019tjn,Valderrama:2019chc,Peng:2019wys} (though predictions of a
$J=\tfrac{1}{2}$, $I=\tfrac{3}{2}$ $\bar{D}^* \Sigma_c$ bound state
have appeared in~\cite{Liu:2019zvb,Yang:2022ezl}).
The two most interesting configurations are probably the $I=\tfrac{1}{2}$,
$\tfrac{3}{2}$, $J=\tfrac{3}{2}$ $\bar{D}^* \Sigma_c^*$ pentaquarks,
for which their isospin quantum numbers are not necessarily clear.
For the central values of the couplings the lower (higher) mass state will be
the octet (decuplet) one, yet it is within the uncertainties of the theory
that the identification might be the opposite one.
Indeed, the calculation of the isospin angle for the lower mass state is
$\theta_I = -(48.0^{+9.9}_{-40.4}) {}^{\circ}$ $({-(52.1-47.1)}^{\circ})$,
which is compatible with $\theta_I < 0$ and the octet identification,
though the large uncertainties do not allow to rule out $\theta_I > 0$.
The relative $J/\psi p$ partial decay width relative to the $P_c(4312)$ is
$3.6^{+0.2}_{-1.7}$ $(3.5-3.6)$, but it could be considerably smaller
if $\theta_I$ turns out to be positive.
The non-observation of this state in the $J/\psi p$ invariant
mass suggests $\theta_I > 0$.

To summarize, we have considered the possibility that the $P_c(4457)$ is a
decuplet $I = \tfrac{3}{2}$ $\bar{D}^* \Sigma_c$ bound state, instead of
the more usual octet $I = \tfrac{1}{2}$ interpretation.
There are a few advantages in explaining the $P_c(4457)$ as a decuplet: 
first, it generates a partial decay width into $J/\psi p$ of the same order
as the $P_c(4312)$ and $P_c(4440)$ (instead of one order of magnitude
larger if it is an octet), which might in turn be in line
with their experimental production fractions.
Second, it implies a smaller $\bar{D}^{(*)} \Lambda_c$ decay width,
which in the octet interpretation turns out to be too large when
compared with the experimental decay width of the $P_c(4457)$.
Third, if the $P_c(4457)$ is a decuplet then the predicted spectrum of
the molecular pentaquarks will be different: seven bound molecular
pentaquarks are predicted in total, but instead of all of them
being octets, there will be a mix of octet and decuplet states.
This might be more compatible with the non-observation of the $J=\tfrac{1}{2}$,
$\tfrac{3}{2}$ $\bar{D}^* \Sigma_c^*$ pentaquarks in the $J/\psi p$ invariant
mass, despite the expectation of them having relatively
large $J/\psi p$ partial decay widths.
The evidence for a decuplet identification of the $P_c(4457)$ is not conclusive
though, as there are a few open issues (production rates of the pentaquarks,
resonance profiles, branching ratios into $J/\psi p$, etc.)
that could tip the balance towards a different
interpretation.

\section*{Acknowledgments}

This work is partly supported by the National Natural Science Foundation
of China under Grants No. 11735003, No. 11835015, No. 11975041, No. 12047503
and No. 12125507, the Chinese Academy of Sciences under Grant No. XDB34030000,
the China Postdoctoral Science Foundation under Grant No. 2022M713229,
the Fundamental Research Funds for the Central Universities and
the Thousand Talents Plan for Young Professionals.
M.P.V. would also like to thank the IJCLab of Orsay, where part of
this work has been done, for its long-term hospitality.

%\bibliography{refs.bib}
%merlin.mbs apsrev4-1.bst 2010-07-25 4.21a (PWD, AO, DPC) hacked
%Control: key (0)
%Control: author (8) initials jnrlst
%Control: editor formatted (1) identically to author
%Control: production of article title (-1) disabled
%Control: page (0) single
%Control: year (1) truncated
%Control: production of eprint (0) enabled
%

\end{document}